\documentclass[aps,preprint]{revtex4}
\usepackage{amsmath}
\usepackage{amssymb}
\usepackage{amsfonts}
\usepackage{graphicx}
\usepackage{subfig}
\usepackage{mathrsfs, mathtools}
\usepackage{pdfpages}
\usepackage{multirow}
\usepackage{float}
\usepackage{color}
\usepackage{xcolor}
\usepackage{caption}
\usepackage{lineno,hyperref}
\hypersetup{colorlinks =true, allcolors = blue}
\usepackage{mathtools}
\renewcommand{\d}{{\rm d}}

\newcommand{\f}{\frac}

\numberwithin{equation}{section}

\providecommand{\U}[1]{\protect\rule{.1in}{.1in}}

\begin{document}
\preprint{HEP/123-qed}
\title{ The static charged black holes with Weyl corrections}
\author{Ahmad Al-Badawi } 
\email{ahmadbadawi@ahu.edu.jo }
\affiliation{Department of Physics, Al-Hussein Bin Talal University 71111,
Ma'an, Jordan.}
\author{\.{I}zzet Sakall{\i}}
\email{izzet.sakalli@emu.edu.tr}
\affiliation{Physics Department, Eastern Mediterranean University, Famagusta 99628, North Cyprus via Mersin 10, Turkey}


\begin{abstract}
 This study investigates static charged black holes with Weyl corrections to the Einstein-Maxwell action. By extending the classical solutions, we incorporate the influence of the Weyl tensor coupling with the electromagnetic field tensor, deriving novel black hole solutions under this framework. The characteristics of the derived spacetimes, including their geometrical structure and physical properties, are meticulously examined. Key features explored include geodesics, photon spheres, and quasinormal modes, providing insights into the stability and dynamics around such black holes. Additionally, the deflection of light and the corresponding shadow cast by these black holes are analyzed, revealing the modifications induced by Weyl corrections. This study aims to enrich the understanding of how higher-order tensor interactions can influence observable astrophysical phenomena in the vicinity of compact objects.

\end{abstract}
\volumeyear{ }
\eid{ }
\date{\today}
\received{}

\maketitle
\tableofcontents
\section{Introduction} \label{sec1}

The exploration of black hole physics is a central theme in modern astrophysics and theoretical physics, providing profound insights into the dynamics of spacetime and gravity under extreme conditions. Traditional models of black holes, primarily grounded in Einstein's General Relativity theory (GR) \cite{isChandra} and supplemented by the Maxwell equations for electromagnetic fields, have been remarkably successful in predicting phenomena such as gravitational lensing \cite{isCunha:2018acu}. However, recent works have proposed various modifications and extensions to these traditional models. For instance, references \cite{referenceA, referenceB, referenceC, referenceD, referenceE, referenceF, referenceG,referenceH,referenceI} have explored different aspects of black hole physics, achieving results that overlap significantly with the goals and calculations presented in this manuscript. It is essential to discuss and compare these works to provide a comprehensive context for the current study. Specifically, Refs. \cite{referenceE} and \cite{referenceF} present models with significant similarities in terms of the coupling of electromagnetic fields with the Weyl tensor that are relevant to our discussion.

Our study focuses on static charged black holes influenced by Weyl tensor corrections \cite{sc1,sc01}. The Weyl tensor \cite{isFrolov:2009qu}, which encapsulates the tidal forces and the intrinsic curvature of spacetime independent of its content, interacts with the electromagnetic field tensor in these modified gravity models. This interaction introduces new dynamics and modifications to the spacetime geometry surrounding black holes. Such theoretical advancements are also crucial for interpreting the increasingly precise data gathered by modern observational technologies such as the Event Horizon Telescope \cite{isEventHorizonTelescope:2019dse} and gravitational wave observatories \cite{isLIGOScientific:2014pky,isHild:2010id,isKAGRA:2013rdx}.

The inclusion of Weyl corrections is motivated by the pursuit to understand and possibly resolve certain theoretical challenges in black hole physics, including the information loss paradox \cite{isPreskill:1992tc,isChen:2014jwq,isGiddings:2006sj,isSakalli:2012zy,isSakalli:2010yy}
 and the singular nature of black hole cores \cite{isZhang:2010vh,isCasadio:2023ymt,isMilosavljevic:2001dd,isMangut:2022ksd,isPourhassan:2023jbs}. Moreover, the corrections are expected to manifest in the observable characteristics of black holes, such as their shadows \cite{isEventHorizonTelescope:2019ths,isFalcke:1999pj,isHioki:2009na,isGralla:2019xty,isUniyal:2022xnq,isOvgun:2018tua}, the bending of light near their horizons, and the spectrum of their quasinormal modes 
\cite{isLIGOScientific:2016lio,isBerti:2009kk,isKonoplya:2011qq,isNollert:1999ji,isSakalli:2022xrb,isSakalli:2011zz,isSakalli:2014wja,isSakalli:2013yha,isSakalli:2022swm,isSakalli:2021dxd,isAl-Badawi:2024iog,isAl-Badawi:2024iax,isAl-Badawi:2024kdw,isAl-Badawi:2024mco}. These observable signatures are essential for the empirical verification of any modifications \cite{isClifton:2011jh} to the GR, thereby providing a more comprehensive understanding of gravitational interactions.

The aforementioned observable physical features of the charged Schwarzschild black hole metric having Weyl corrections (SCWBH) \cite{sc1} considered in this paper are examined through some analytical methods. One primary area of focus is the modification of geodesic paths \cite{isCarter:1968rr}, which influences how particles and light propagate in the vicinity of these black holes. Another area explored is the stability of these spacetimes, assessed through the analysis of quasinormal modes, which reflect the inherent "ringing" of spacetime due to perturbations \cite{isChandrasekhar:1988kt}. Furthermore, this research investigates the optical effects linked to black holes, particularly focusing on the properties of photon spheres and the intricate formation of black hole shadows, which can be detected using modern astronomical tools.

The organization of the paper is as follows: Sec. \ref{sec2} provides a brief review of the theoretical framework used for incorporating Weyl tensor corrections into the Einstein-Maxwell equations and serves the fundamental equations describing the modified black hole solutions. Section \ref{sec3} analyzes the geodesic structure in the SCWBH spacetime, highlighting changes in both particle and photon orbits. Section \ref{sec4} discusses the properties of photon spheres and the visual appearance of the black hole shadows, predicting alterations in observable features due to Weyl corrections. Section \ref{sec5} focuses on the quasinormal modes of the SCWBH, exploring implications for their dynamical stability and response to external perturbations. Section \ref{sec6} examines the deflection of light by these modified black holes using the Gauss-Bonnet theorem (GsBnTh) to calculate bending angles under a weak field approximation. In conclusion, Sec. \ref{sec7} integrates the results, examines their significance for both theoretical physics and observational astronomy, and proposes possible future research studies in this dynamic area.

\section{The static charged black holes with Weyl corrections} \label{sec2}

 In this section, we consider the static SCWBH \cite{sc1} and provide some of its physical characteristics. According to \cite{sc01, sc1}, the coupling of an electromagnetic field with a Weyl tensor is represented by the following action:
\begin{equation}
S=\int \mathrm{d}^{4}x\sqrt{-g}\left( R-\frac{1}{4}F^{\mu \nu }F_{\mu \nu }+\alpha C^{\mu \nu \rho \sigma}
F_{\mu \nu } F_{\rho\sigma}\right),\label{action1}
\end{equation}
where $\alpha$ is a coupling constant, $C^{\mu \nu \rho \sigma}$ is called Weyl tensor, $g$ represents the determinant of the metric tensor, $R$ denotes the Ricci scalar, $F_{\mu u }=\partial_\mu A_u-\partial_u A_\mu$ stands for the electromagnetic tensor is the Weyl tensor. This action was first introduced in \cite{referenceH}, which provides a foundational basis for our study. The motivation for this equation lies in its ability to capture the interaction between electromagnetic fields and spacetime curvature, extending beyond the traditional Maxwell framework. The vector potential $A_\mu$ is given by
\begin{equation}
    A_\mu=\left(\phi(r),0,0,0\right). \label{vp1}
\end{equation}
For $\alpha = 0$, Eq. (\ref{action1}) simplifies to the standard Einstein-Maxwell action. Consequently, the SCWBH exhibits a spherically symmetric spacetime \cite{sc1}, described by
\begin{equation}
\text{d}s^{2}=-f(r) \text{d} t^{2}+\frac{1}{f(r)} \text{d}r^{2}+g(r)(\text{d}\theta ^{2}+\sin^{2}\theta \text{d}\varphi
^{2}) , \label{M1}
\end{equation}
where
\begin{equation}
f(r)=1-\frac{2M}{r}+\frac{q^2}{r^2}-\frac{4\alpha q^2}{3r^4}\left(1-\frac{10M}{3r}+\frac{26 q^2}{15r^2}\right),
\label{equ2}
\end{equation}
\begin{equation} \label{egu}
    g(r)=r^{2}+\frac{4\alpha q^2}{9r^2},
\end{equation}
\begin{equation}
 \phi(r)= \frac{q}{r}+\frac{\alpha q}{r^3}\left(\frac{M}{r}-\frac{37 q^2}{45r^2}\right),  
\end{equation}
where $M$, $q$, and $\alpha$ represent black hole mass, electric charge, and coupling parameter, respectively. The event horizon, \( r_+ \), is the largest real root of the equation \( f(r) = 0 \). This horizon radius is crucial for further calculations related to the thermodynamic properties of the black hole.

The surface gravity \( \kappa \) at the horizon is computed using:
\begin{equation}
    \kappa = \frac{1}{2} \frac{\text{d}f}{\text{d}r} \bigg|_{r = r_+},
\end{equation}
where
\begin{align}
    \frac{\text{d}f}{\text{d}r} \bigg|_{r = r_+} &= \frac{2M}{{r_+}^2} - \frac{2q^2}{{r_+}^3} + \frac{16\alpha q^2}{3{r_+}^5} \left(1 - \frac{10M}{3{r_+}} + \frac{26q^2}{15{r_+}^2}\right) - \frac{4\alpha q^2}{3{r_+}^4} \left(\frac{10M}{3{r_+}^2} - \frac{52q^2}{15{r_+}^3}\right).
\end{align}

On the other hand, due to the structure of $f(r)$, an explicit analytical solution for $r_{+}$ is not available; this calculation can be performed numerically or using further assumptions about the dominant terms in $f(r)$. Here, one can estimate $\kappa$ assuming $r_{+}$, which is the Schwarzschild radius for simplicity, and hence it is a rough approximation ignoring charge and scalar field corrections. Thus, surface gravity $\kappa$ can be approximated to
\begin{align}
\kappa & \approx \frac{1}{720M^7} \bigg(32400M^{12} - 32400M^{10}q^2 + 900M^8q^2(-26\alpha + 9q^2) + 25740M^6\alpha q^4 \\ \notag
&+65M^4\alpha q^4(65\alpha - 108q^2) - 5070M^2\alpha^2q^6+1521\alpha^2q^8\bigg)^{\frac{1}{2}}.
\end{align}

This expression encapsulates the gravitational effects near the horizon, significantly influenced by the black hole's characteristics. The above calculation demonstrates the complex interaction between gravitational forces and quantum field effects near the horizon of a black hole. In fact, the Unruh acceleration \cite{isUnruh:1976db}, correlating with the surface gravity, highlights the infinite acceleration required to maintain a stationary position at the event horizon, as predicted by GR. This forms a critical connection with the observed Hawking radiation and the thermal properties associated with black hole horizons \cite{isCrispino:2007eb}.

The Hawking temperature $T_H$ of the black hole is related to the surface gravity by \cite{isWald:1984rg}:
\begin{equation}
    T_H = \frac{\kappa}{2\pi}.
\end{equation}

The entropy $S$ of the black hole \cite{isWald:1984rg}, following the area law, is:
\begin{equation}
    S_{BH} = \frac{A_H}{4},
\end{equation}
where $A_H$ is the area of the event horizon given by:
\begin{equation}
    A_H= 4\pi \left(r_+^2 + \frac{4\alpha q^2}{9r_+^2}\right).
\end{equation}

\section{Geodesics}\label{sec3}

\subsection{Null geodesics in SCWBHs}
{\color{black}To analyze the geodesics of massless particles in the spacetime described by Eq. (\ref{M1}), we use the Lagrangian and Hamilton-Jacobi equations, following the methodology outlined in Chandrasekhar's textbook \cite{isChandrasekhar:1988kt}. The Euler-Lagrange equation is expressed as
\begin{equation}
\frac{\text{d}}{\text{d} s}\left( \frac{\partial \mathscr{L}}{\partial \overset{\cdot }{x^{\mu }}}\right) =\frac{\partial \mathscr{L}}{\partial x^{\mu }},
\end{equation}
where $s$ denotes the affine parameter of the light's trajectory, a dot indicates differentiation with respect to $s$, and $\overset{\cdot }{x^{\mu }}$ represents the four-velocity of the light ray. The Lagrangian $\mathscr{L}$ is given by
\begin{equation}
 \mathscr{L}=-f(r)\dot{t}^2+\frac{\dot{r}^2}{f(r)}+g(r) \left(\dot{\theta}^2 + \sin^2\theta \dot{\varphi}^2\right). \label{lagran}
\end{equation}
Given the static and spherically symmetric nature of spacetime, we restrict our analysis to the equatorial plane $(\theta=\pi/2)$ without loss of generality. By setting $\theta=\pi/2$ and $\dot{\theta}=0$, we derive the Euler-Lagrange equations for the $t$ and $\varphi$ coordinates:
\begin{equation}
    f(r)\dot{t}=E, \label{en9}
\end{equation}
\begin{equation}
   g(r) \dot{\varphi}=L,\label{ang8} 
\end{equation}
where $E$ and $L$ are the conserved quantities of energy and angular momentum along the geodesics, respectively. Using these equations, we can express the Lagrangian (\ref{lagran}) as
\begin{equation}
    E^2=\dot{r}^2+V_{\text{eff}},\label{en19}
\end{equation}
where the effective potential is
\begin{equation}
    V_{\text{eff}}=\frac{L^2}{g(r)}f(r). \label{effp} 
\end{equation}
To explore the trajectory of light rays, we derive the relationship between $r$ and $\varphi$, expressed as
\begin{equation}
   \left( \frac{\text{d} r}{\text{d} \varphi}\right)^2=\frac{g^2(r)}{b^2}-g(r)f(r), \label{phi1}
\end{equation}
where the impact parameter $b$ is defined as $b = \frac{L}{E}$. Subsequent subsections will separately investigate the radial null geodesics and null geodesics with angular momentum.

\subsection{Radial null geodesics $(L = 0)$
}
{\color{black}For radial motion with zero angular momentum, the effective potential vanishes. Hence, using Eqs. (\ref{en9}) and (\ref{ang8}), we can derive the equations that describe the relationship between coordinate time $t$, radial distance $r$, and the affine parameter $s$:
\begin{equation}
    \frac{\text{d}t}{\text{d}r}=\pm\frac{1}{f(r)},
\end{equation}
\begin{equation}
    \frac{\text{d}s}{\text{d}r}=\pm\frac{1}{E},
\end{equation}
where the $\pm$ signs indicate outgoing and ingoing trajectories, respectively. Figure \ref{fig1a} illustrates the time behavior of incoming photons in the spacetime of SCWBH. Assuming photons start at $r = r_0$ when $t=0$ and approach $r = r_+$, the figure shows that photons reach the horizon after an infinite period, consistent with the behavior observed in the Schwarzschild scenario \cite{isChandrasekhar:1988kt}. 
\begin{figure}
    \centering
    \includegraphics{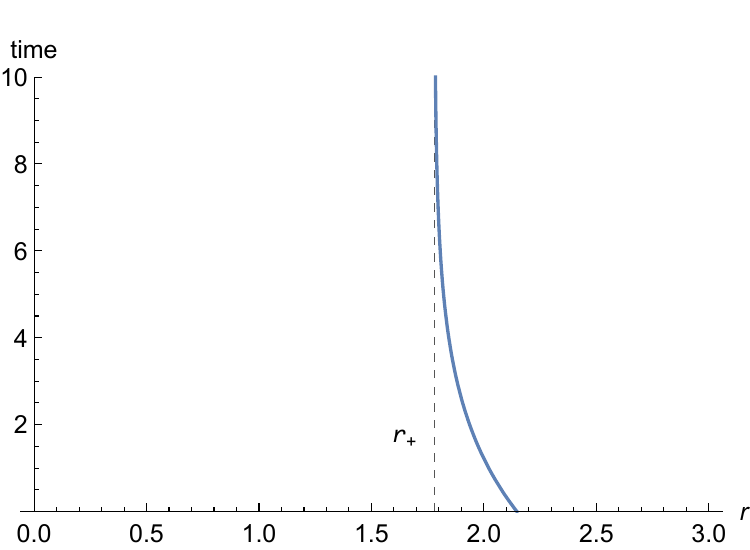}
    \caption{The behavior of the time coordinate $t$ as a function of radius $r$. Here, $M=1, \alpha=0.3$, $q=0.6$, and $r_+=1.7806$}
    \label{fig1a}
\end{figure}}

\subsection{Geodesics with angular momentum $L\neq 0$}

{\color{black}The effective potential for the angular motion of photons around a SCWBH is given by:
\begin{equation}
   V_{\text{eff}}=\frac{9r^2 L^2}{9r^4+4\alpha q^2}\left(1-\frac{2M}{r}+\frac{q^2}{r^2}-\frac{4\alpha q^2}{3r^4}\left(1-\frac{10M}{3r}+\frac{26 q^2}{15r^2}\right)\right). \label{effp88}
\end{equation}
For non-radial geodesics, setting $L = 1$ simplifies our calculations, allowing us to use $b = 1/E$. Figure \ref{fig2a} displays a plot of the effective potential as it varies with radial distance and energy levels. 
\begin{figure}
    \centering
    \includegraphics{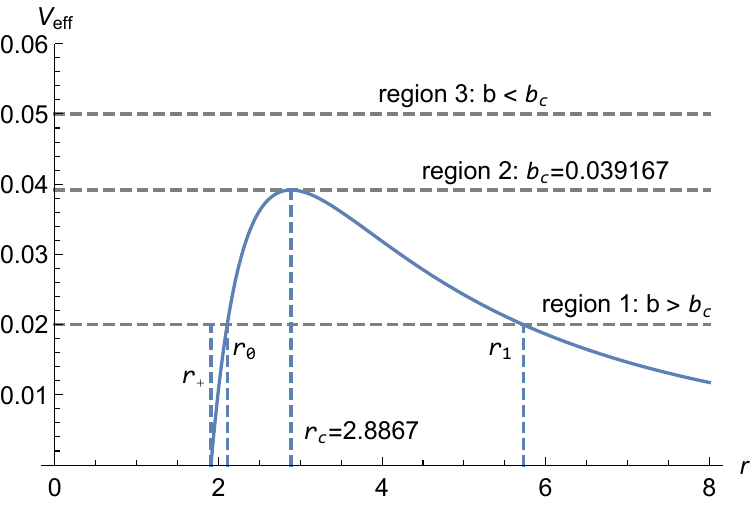}
    \caption{The effective potential of an SCWBH and various corresponding energy levels. Here, $\alpha=0.4$, $q=0.4$, and $M=1$ are set.}
    \label{fig2a}
\end{figure}
Equation (\ref{en19}) shows that photon motion is significantly influenced by energy levels. To demonstrate the dynamics of various photon motions, we plotted the effective potential (\ref{effp88}) alongside some representative energy levels in Figure \ref{fig2a}. Based on the inward motion of light rays, we categorize their movements as follows:

1. When $b > b_c$ (region 1), light rays originating from $r \geq r_1$ meet a potential barrier and are repelled back towards $r = r_1$. However, if the photon begins its journey where $r_+<r \leq r_0$, it crosses the event horizon.

2. When $b = b_c$ (region 2), photon trajectories initiate inward from $r > r_c$ and asymptotically approach the point $r = r_c$, requiring an infinite affine time to reach it. This implies that the photons never actually reach this point but rather continue to orbit around the photon sphere indefinitely.

3. When $b < b_c$ (region 3), photons start at infinity and, in the absence of a potential barrier, directly fall into the black hole.}
   
\subsection{ Circular orbits}
When the effective potential reaches its maximum, photons can move in circular orbits. Nonetheless, these circular orbits are unstable because even minor changes in impact parameter $b$ cause the photon to diverge from the peak and drift away. The positions of the circular photon orbits can be inferred by the following conditions:
\begin{equation}
  V_{eff}=\frac{1}{b_c^2}, \hspace{1cm} \frac{\text{d} V_{eff}}{\text{d}r} =0 .
\end{equation}
The ISCO orbit indicates the minimum radius where circular motion remains stable. Our goal is to determine the ISCO radius of the massive particle in SCWBH using the following standard conditions:\begin{equation}
V_{eff}=0,\hspace{1cm}V_{eff}^{\prime }=0,\hspace{1cm}V_{eff}^{\prime \prime }\geq 0,  \label{veff44}
\end{equation}%
where $V_{eff}$ is given by Eq. (\ref{effp88}). 
Based on the three conditions above, the ISCO radius equation can be derived as follows: \begin{equation}
\left.\left(2f'gg'-\frac{ff''gg'}{f'}+fgg''-2f(g')^2 \right) \right\vert _{r=r_{ISCO}}=0. \label{isco}
 \end{equation}
 It is obvious that equation (\ref{isco}) is extremely difficult to solve analytically. As a result, we solve it numerically and provide our findings in Table \ref{tablea2}.  The Table shows that for a fixed $\alpha$, increasing the charge parameter $q$ reduces the size of the ISCO radius. While the charge parameter $q$ remains constant, increasing the coupling parameter causes the ISCO radius to slightly rise. As a result, we can deduce that the charge parameter has a greater impact on the size of the ISCO than the coupling parameter does. Note we create a graphic (Fig. \ref{figisc}) to see the impact of $q$ and $\alpha$ on the ISCO radius for massive particles.
 \begin{center}
\begin{tabular}{|c|c|c|c|c|c|c|}  \hline 
$ISCO$ & $\alpha =-0.2$ & $-0.1$ & $0$ & $0.1$ & $0.2$ \\ \hline
$q=0.5$ & $5.60418$ & $5.60541$ & $5.60664$ & $5.60788$ & $5.60911$ \\ \hline
$0.6$ & $5.41616$ & $5.418$ & $5.41984$ & $5.42169$ & $5.42354$ \\ \hline
$0.7$ & $5.17999$ & $5.1826$ & $5.18523$ & $5.18786$ & $5.19049$ \\ \hline
$0.8$ & $4.88364$ & $4.8872$ & $4.89077$ & $4.89435$ & $4.89795$ \\ \hline
$0.9$ & $4.50463$ & $4.50917$ & $4.51375$ & $4.51836$ & $4.52302$ \\ \hline
$1$ & $3.99078$ & $3.99533$ & $4.$ & $4.00478$ & $4.00967$ \\ \hline  
\end{tabular}\captionof{table}{The numerical values of $ISCO$ orbit for
different values of $q$ and $\alpha$
parameters when $M = 1$.} \label{tablea2}
 \end{center}
 
 \begin{figure}
     \centering
     \includegraphics{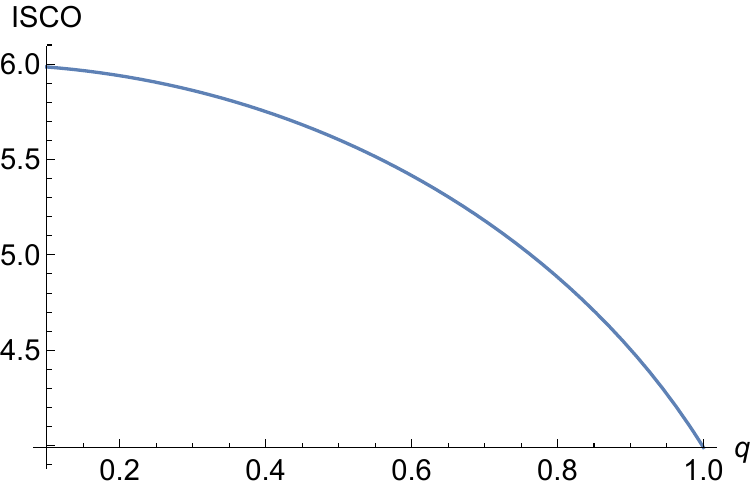}
     \caption{Graph of ISCO radius versus $q$ for $\alpha=-0.2$.}
     \label{figisc}
 \end{figure}
\subsection{Geodesics in terms of the variable $u=1/r$}

In this part, we will look at how photons move near the SCWBH. In general, to analyse the motion of a test particle in the gravitational field, we are interested in a particle track in the $r-\varphi $ plane.
Hence, It is convenient to use the variable $u=1/r$. Using Eq. (\ref{phi1}) the trajectory equation that governing this motion is 
\begin{equation}
 \frac{\text{d}u}{\text{d} \varphi } =\sqrt{\left( 1+\frac{4\alpha q^2 u^4}{9}\right)\left( 2Mu-q^2 u^2+\frac{4}{45}\alpha q^2 u^4(15-50Mu+26 q^2 u^2) +\frac{1+\frac{4\alpha  q^2 u^4}{9}}{bu^2} -1\right)}.\label{gu5}
\end{equation}%
It should be noted that Eq. (\ref{gu5}) provides the Reissner-Nordstr\"{o}m equation of motion for photon in spacetime when $\alpha=0$ and the Schwarzschild equation for $q=0$.\\ As discussed in the preceding section, the motion of a photon is determined by its energy level (Figure \ref{fig2a}). When $b > b_c$ photons deflect at $u_1 = 1/r_1$, which is the root of the Eq. (\ref{gu5}). Therefore, photon reaches the turning point $u = u_1 $ (or $r = r_1$) and is reflected, resulting in an unbounded orbit. When $b = b_c$, photons form an unstable circular orbit with radius $r = r_c$, circling the black hole on the photon sphere. Finally, for $ b < b_c$, light beams continue to go inward until they are captured by the black hole.\\ It is clear that the above equations have no analytical solutions, hence a numerical analysis is required to investigate the geodesic paths of massive particles. Figure \ref{figa4} illustrates the numerically plotted photon paths resulting from Eq. (\ref{gu5}). The Figure \ref{figa4} shows the photon trajectories near SCWBH in Euclidean polar coordinates ($r,\varphi $). The black disc in the Figure represents the region inside the black hole's event horizon. It is encircled by a dashed green curve that depicts the circular photon orbit while the red dotted correspond to the shadows.
\begin{figure}
    \centering
    \includegraphics{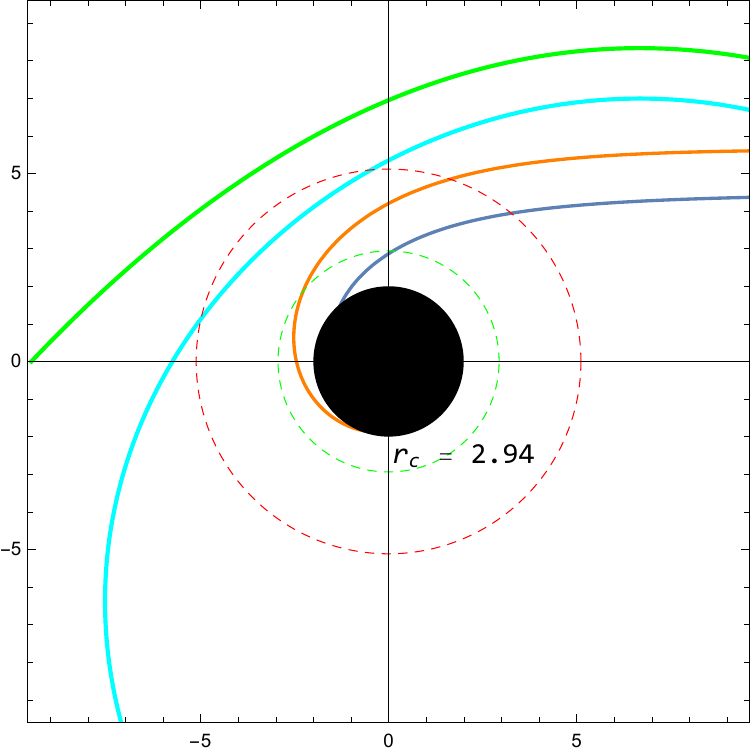}
    \caption{We show four light trajectories with a mass $M = 1$, $q=0.3, \alpha=-0.5$ and an impact factor of 4 as initial conditions. Photon spheres are represented by green dotted lines, while shadows of the black hole are represented by red dotted lines.}
    \label{figa4}
\end{figure}

\section{Shadow OF SCWBH} \label{sec4}

An observer placed far away from the black hole can perceive the dispersion and unstable circular motion. Specifically, the unstable photon circular orbit acts as a border between the scattering and falling trajectories. This black hole image shows a photon ring enclosing the black region which is known as the black hole shadow. In order to study shadow observational feature, it is necessary to consider the null curves around a static charged black hole with Weyl corrections (\ref{M1}). Geodesic equations can be obtained by using Hamilton-Jacobi equations as follows: 
\begin{equation}
\frac{\partial \mathcal{S}}{\partial \sigma }=-\frac{1}{2}g^{\mu \nu }\frac{%
\partial \mathcal{S}}{\partial x^{\mu }}\frac{\partial \mathcal{S}}{\partial
x^{\nu }},
\end{equation}%
where $\mathcal{S}$ is the Jacobi action. The following Jacobi action separable solution reads
\begin{equation}
\mathcal{S}=-Et+L \phi +\mathcal{S}_{r}\left( r\right) +\mathcal{S}%
_{\theta }\left( \theta \right) ,
\end{equation}%
where $E$ and $\ell $ are the two Killing vectors of metric (\ref{M1}), given by%
\begin{equation}
E=-f(r) \dot{t}  \label{E1}
\end{equation}%
\begin{equation}
L =g(r) \sin ^{2}\theta \overset{\cdot 
}{\phi }\text{.}  \label{An1}
\end{equation}%
Therefore, the geodesic equations read
\begin{equation}
\frac{\text{d}t}{\text{d}\sigma }=\frac{E}{f },\qquad \frac{\text{d}\phi }{\text{d}\sigma 
}=-\frac{L }{g\sin ^{2}\theta },
\end{equation}%
\begin{equation}
r^2 \frac{\text{d}r}{\text{d}\sigma }=\pm \sqrt{\mathcal{R}\left( r\right) },\qquad r^2 \frac{\text{d}\theta }{\text{d}\sigma }=\pm \sqrt{\Theta \left( \theta \right) },  \label{R1}
\end{equation}%
where $\mathcal{K}$ is the Carter separation constant and 
\begin{equation}
\mathcal{R}\left( r\right) =r^{4}E^{2}-\left(\delta g+ \mathcal{K}+L ^{2}\right)
gf ,  \label{R2}
\end{equation}%
\begin{equation}
\Theta \left( \theta \right) =\mathcal{K}-L ^{2}\cot^2 \theta .
\end{equation}%
We introduce dimensional quantities called impact parameters as follows: 
\begin{equation}
\eta =\frac{\mathcal{K}}{E^{2}},\qquad \text{ }\zeta =\frac{L }{E}.
\label{impact1}
\end{equation}%
We are interested in spherical light geodesics constrained to a constant coordinate sphere
radius $r$ with $\overset{\cdot }{r}=0$ and $\overset{\cdot \cdot }{r}=0$ also known as spherical photon orbits. It is recommended that photons in circular orbits satisfy the maximum effective potential conditions:
\begin{equation}
V_{eff}\left( r\right) \left\vert _{r=r_{ps}}\right. =0,\hspace{1cm} V_{eff}^{\prime
}\left( r\right) \left\vert _{r=r_{ps}}\right. =0,
\end{equation}%
or,%
\begin{equation}
\mathcal{R}\left( r\right) \left\vert _{r=r_{ps}}\right. =0, \hspace{1cm} \mathcal{R}%
_{eff}^{\prime }\left( r\right) \left\vert _{r=r_{ps}}\right. =0.  \label{R3}
\end{equation}%
where $r_{ps}$ is the photon sphere and marks the location of the apparent
image of the photon rings. If we consider metric (\ref{M1}) then we can write
the radius of the photon sphere as the solution of the equation 
\begin{equation}
g^{\prime }\left( r_{ps}\right) f\left( r_{ps}\right) -g\left( r_{ps}\right)
f^{\prime }\left( r_{ps}\right) =0,
\end{equation}%
or explicitly%
\begin{equation*}
405(3M-r)r^9+832q^6\alpha^2-90q^2 r^5(9r^3+72M\alpha-20\alpha r)+
\end{equation*}
\begin{equation}
48q^4\alpha r(78r^3-25M\alpha+5\alpha r) =0. \label{rps1}
\end{equation}
Moreover, let us examine the observable, namely the shadow radius $R_{sh}$, which is responsible for providing significant information about the black hole shadow:
\begin{equation}
R_{sh}=\left. \sqrt{\frac{g\left( r_{ps}\right) }{f\left( r_{ps}\right) }}%
\right\vert _{r=r_{ps}}, \label{rshadow}
\end{equation}%
and thus it coincides in value with the impact parameter itself.\\
Equation (\ref{rps1}) is quite difficult to solve analytically. As a result, we provide numerical analysis presented in Table \ref{tablea1} and graphs as that show how the coupling parameter $\alpha$ affects the photon radius of mass particles as well as the radius of the black hole shadow. 
\begin{center}
\begin{tabular}{|c|c c |c c|c c|c c|}  \hline 
$M=1$& $q=0.3$ &  & $q=0.5$ &  & $q=0.7$ &  & $q=0.9$ &  \\ \hline
$\alpha $ &$r_{ps}$ & $R_{sh}$  &  $r_{ps}$ & $R_{sh}$ & $r_{ps}$ & $R_{sh}$ & $r_{ps}$ & $R_{sh}$   \\ \hline
$-0.5$ & $2.94038$ & $5.11683$ & $2.82832$ & $4.96827$ & $2.64181$ & $4.72270
$ & $2.33994$ & $4.33071$ \\ 
$-0.4$ & $2.94005$ & $5.11682$ & $2.82724$ & $4.96820$ & $2.63895$ & $4.72233
$ & $2.33171$ & $4.32874$ \\ 
$-0.3$ & $2.93973$ & $5.11682$ & $2.82616$ & $4.96813$ & $2.63603$ & $4.72194
$ & $2.32304$ & $4.32662$ \\ 
$-0.2$ & $2.9394$ & $5.11681$ & $2.82507$ & $4.96806$ & $2.63306$ & $4.72154$
& $2.31386$ & $4.32434$ \\ 
$-0.1$ & $2.93908$ & $5.11680$ & $2.82398$ & $4.96799$ & $2.63003$ & $4.72112
$ & $2.30412$ & $4.32188$ \\ 
$0$ & $2.93875$ & $5.11679$ & $2.82288$ & $4.96791$ & $2.62694$ & $4.72068$
& $2.29373$ & $4.31923$ \\ 
$0.1$ & $2.93842$ & $5.11679$ & $2.82176$ & $4.96784$ & $2.62379$ & $4.72023$
& $2.28259$ & $4.31635$ \\ 
$0.2$ & $2.93809$ & $5.11678$ & $2.82065$ & $4.96776$ & $2.62058$ & $4.71975$
& $2.27058$ & $4.31323$ \\ 
$0.3$ & $2.93776$ & $5.11677$ & $2.81952$ & $4.96768$ & $2.61730$ & $4.71926$
& $2.25754$ & $4.31699$ \\ 
$0.4$ & $2.93743$ & $5.11676$ & $2.81838$ & $4.96759$ & $2.61395$ & $4.71875$
& $2.24324$ & $4.30608$ \\ 
$0.5$ & $2.93710$ & $5.11675$ & $2.81724$ & $4.96751$ & $2.61052$ & $4.71822$
& $2.22738$ & $4.30196$ \\ \hline  
\end{tabular}
\captionof{table}{The numerical values of $r_{ps}$ and $R_{sh}$.} \label{tablea1}
\end{center}
Table \ref{tablea1} shows that both radii ($r_{ps}$ and $R_{sh}$) drop as the coupling parameter and charge rise. The coupling parameter causes a very tiny change. As a result, charge has a greater impact on both radii than the coupling parameter.  When the charged black hole with Weyl correction is compared to the Reissner-Nordstr\"{o}m (RN) photon sphere and the shadow radius, it turns out that when the coupling parameter is negative, both radii are bigger than the RN one, however when the coupling parameter is positive, the radii are less than the RN. 
\\Figure \ref{shadow12} 
depicts the variations of both photon sphere $r_{ps}$ and the shadow radius $R_{sh}$  
for the static charged black holes with Weyl corrections.
\begin{figure}
    \centering
    \includegraphics[scale=0.6]{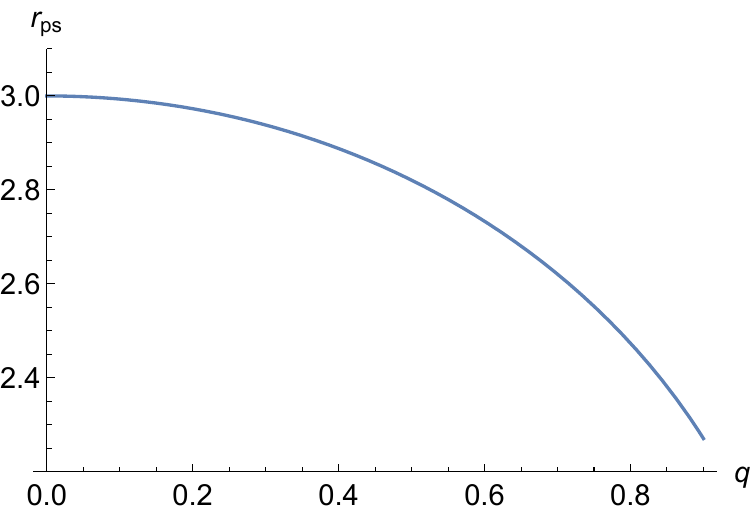}
\includegraphics[scale=0.6]{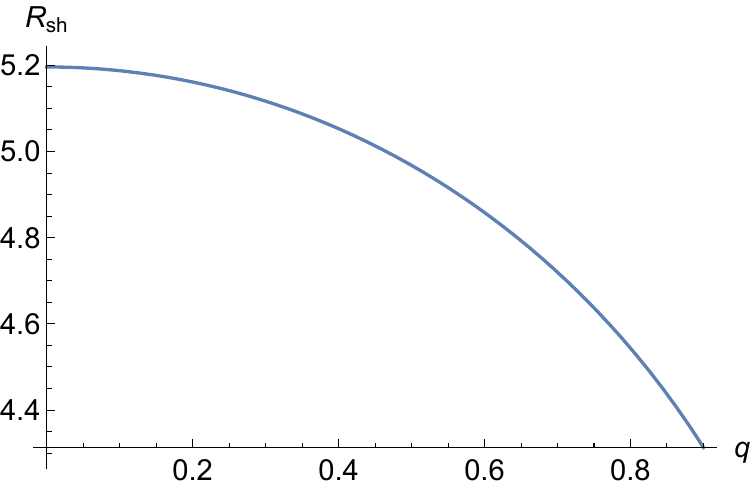}
    \caption{ The variations of the $r_{ps}$ (left) and $R_{sh}$ (right) versus $q$ for coupling parameter $\alpha=0.2$.}
    \label{shadow12}
\end{figure}
A plot of the SCWBH shadow boundary sizes for different values of electric charge $q$ is shown in Figure \ref{sh21} with $\alpha =- 0.5$ and $0.5$, respectively.
\begin{figure}
    \centering
    \includegraphics[scale=0.6]{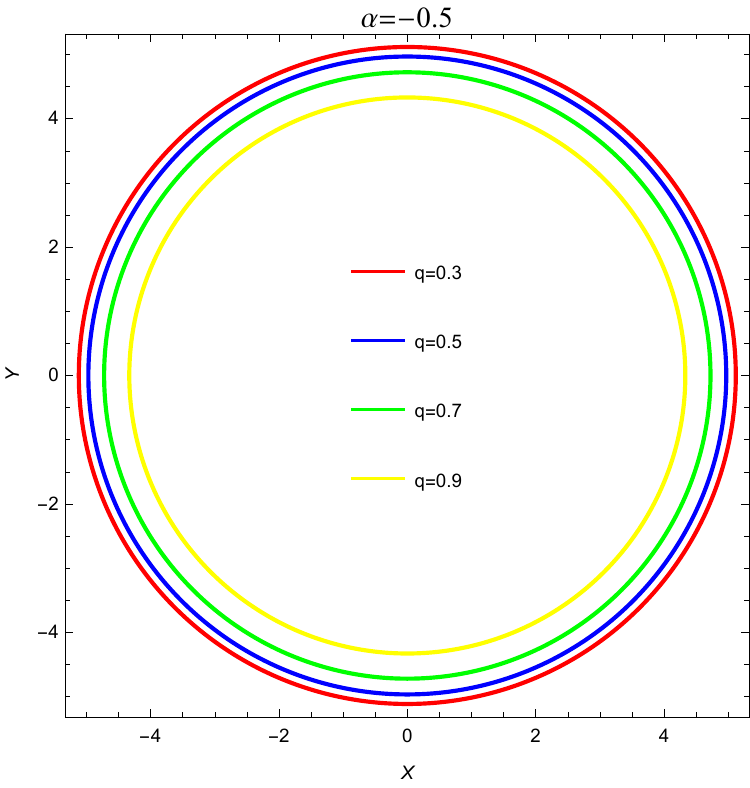}
\includegraphics[scale=0.6]{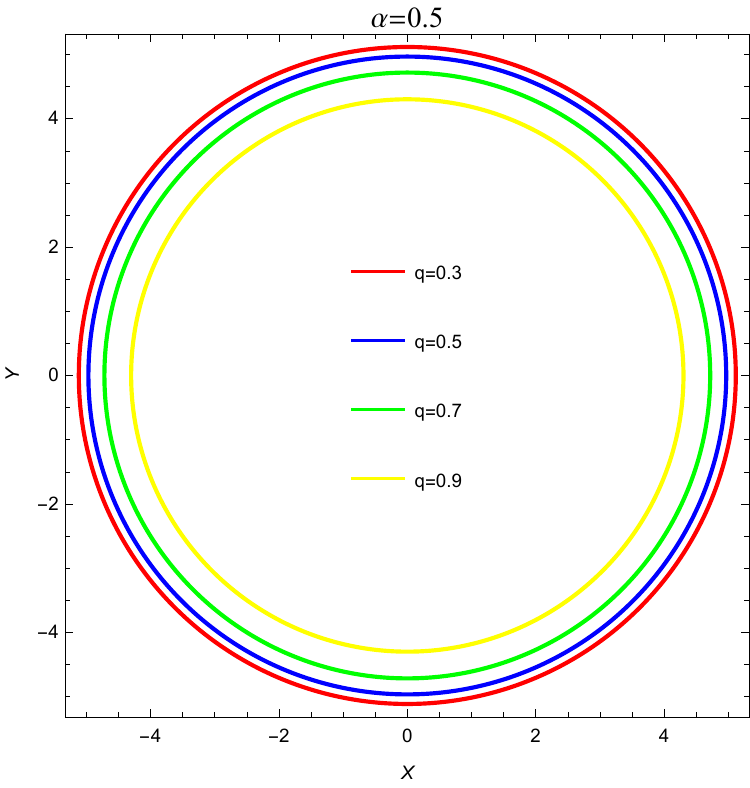}
    \caption{ The shadow of the SCWBH with several values of electric charge $q$.}
    \label{sh21}
\end{figure}

\section{QNM\lowercase{s} of SCWBH} \label{sec5}
In this section, we explore the QNMs of SCWBH by examining spin-$0$ particles with a massive scalar field denoted by $\Psi$. Here, it is more straightforward to express the Klein-Gordon equation in a curved spacetime context as follows:
\begin{equation}\label{iseqspin0}
\big(\square+m_\psi^2\big)\Psi=\f{1}{\sqrt{-g}}\partial_a\big(g^{ab}\sqrt{-g}\,\partial_b\Psi\big)+m_\psi^2\Psi=0,\quad \sqrt{-g}=g(r)sin\theta,
\end{equation}
where $m_\psi$ is the mass of the field. Setting the prime symbol as a radial derivative, the massive spin-$0$ field satisfies the equation of motion \eqref{iseqspin0}. Using the metric \eqref{M1}, this becomes
\begin{equation}
	-\partial_{t}^2\Psi+\f{f(r)}{g(r)}\left(\f{1}{\sin\theta}\partial_\theta(\sin\theta\,\partial_\theta)+\csc^2\theta\,\partial_\varphi^2\right)\Psi+\f{f(r)}{g(r)}\bigg(f(r)g(r)\Psi'\bigg)'+f(r)m^2_\psi\Psi=0\,.
\end{equation}
Using the following ansatz:
\begin{equation}\label{iseqanzatz}
	\Psi=\Phi_0(r)S^0_{\ell,m}(\theta,\varphi)e^{-i\omega t},
\end{equation}
where $\ell$ and $m$ represent the orbital and magnetic quantum numbers, respectively. Here, $S^0_{\ell,m}$ are the spin-$0$ weighted spherical harmonics for the angular modes $\ell,m$, satisfying the following angular equation:
\begin{equation} \label{ismkge}
	\left(\f{1}{\sin\theta}\partial_\theta(\sin\theta\,\partial_\theta)+\csc^2\theta\,\partial_\varphi^2++\lambda_\ell^0\right)S_{\ell,m}^0=0,
\end{equation}
where the separation constant, the so-called eigenvalue, is $\lambda_\ell^0\equiv\ell(\ell+1)$. Thus, the radial part can be decoupled and one gets
\begin{equation}\label{iskge}
\left(\omega^2+f(r)m_\psi^2-\lambda_\ell^0\f{f(r)}{g(r)}\right)\Phi_0+\f{f(r)}{g(r)}\bigg(f(r)g(r)\Phi_0'\bigg)'=0.
\end{equation}
In short, Eq. \eqref{iskge}  is nothing but the radial Klein-Gordon equation for a massive spin-0 field. For all of this section, let us define a generalized tortoise coordinate $r^*$ as 
\begin{equation}\label{istortoise}
	\f{\d r^*}{\d r}=\f{1}{f(r)}.
\end{equation}
To derive the expression of the potential for spin-0 field, let us consider the general redefinition of the wave function as
\begin{equation}\label{istransformation}
	Z_0\equiv \Phi_0(r)\sqrt{g(r)}\,,
\end{equation}
where all quantities are functions of $r$. Now, we can find a wave function $Z_0$ satisfying the general Schr\"odinger-like equation:
\begin{equation} \label{isZeq}
\partial_*^2Z_0+\Big(\omega^2-V_0\big(r(r^*)\big)\Big)Z_0=0, 
\end{equation}
with potential $V_0$, and where $\partial_*$ denotes the derivative with respect to the tortoise coordinate $r^*$. For the massive spin-0 field, there is no complex term in Eq. \eqref{iseqspin0}, and all the terms are already decreasing faster than $1/r^2$ at spatial infinity because of the fall-offs:
\begin{equation}\label{falloffs}
	f(r)\underset{r\rightarrow+\infty}{\longrightarrow}1 ,\quad g(r)\underset{r\rightarrow+\infty}{\sim}r^2\,.
\end{equation}
This means that at spatial infinity the functions $f(r)$ and $g(r)$ satisfy the asymptotic conditions. By implementing the transformations \eqref{istortoise} and \eqref{istransformation} on Eq. \eqref{iskge}, we derive a one-dimensional Schr\"odinger like wave equation, represented as Eq. \eqref{isZeq}, characterized by a potential denoted by 
\begin{align}
	V_0\big(r(r^*)\big)&=-f(r)m_\psi^2+\ell(\ell+1)\f{f(r)}{g(r)}+\f{1}{2}\sqrt{\f{f(r)^2}{g(r)}}\left(\sqrt{\f{f(r)^2}{g(r)}}g'(r)\right)' , \nonumber \label{effQ1} \\
	&=-f(r)m_\psi^2+\ell(\ell+1)\f{f(r)}{g(r)}+\f{\partial_*^2\sqrt{g(r)}}{\sqrt{g(r)}}.
\end{align} 
This is the potential for the massive spin-0 field in the SCWBH metric \eqref{M1}.\\ 
In the  literature, a variety of strategies for obtaining analytical solutions for quasinormal modes. Will and Iyer \cite{will1,iyer} proposed one of the most well-known techniques, the WKB method. Later, Konoplya improved the approach to the sixth order \cite{konop}. In our computations, we take into account scalar field perturbations. To explore the massless scalar QNMs of the SCWBH, we consider the effective potential (\ref{effQ1}) but by setting $m_\psi=0$. The effect of both the electric charge and the coupling parameter on the QNMs is investigated. \\ In Tables \ref{tqnm1} and \ref{tqnm2}, we present the calculated quasinormal frequencies of
the SCWBH using the following model parameters: $M = 1,  l = 2,$ and overtone number $ n = 0$. 

\begin{center}
\begin{tabular}{|c|c|c|c|} \hline
& $\alpha =-0.3$ & $\alpha =-0.2$ & $\alpha =-0.1$  \\ \hline
$q=0.3$ & $0.53870-0.089234i$ & $0.53879-0.089303i$ & $0.536271-0.089123i$ \\ 
$0.4$ & $0.54534-0.089921i$ & $0.54531-0.089893i$ & $0.545281-0.089865i$  \\ 
$0.5$ & $0.55429-0.090710i$ & $0.55423-0.090659i$ & $0.554184-0.090607i$  \\ 
$0.6$ & $0.56619-0.091688i$ & $0.56610-0.091597i$ & $0.566024-0.091505i$  \\ 
$0.7$ & $0.58192-0.092849i$ & $0.58179-0.092684i$ & $0.581668-0.092517i$  \\ 
$0.8$ & $0.60298-0.094143i$ & $0.60281-0.093829i$ & $0.602617-0.093506i$  \\ 
$0.9$ & $0.63228-0.095374i$ & $0.63205-0.094702i$ & $0.631765-0.093985i$  \\ 
$1$ & $0.67658-0.095610i$ & $0.67642-0.09374i$ & $0.675948-0.091486i$  \\ \hline
\end{tabular} 
\captionof{table}{Variation of $l = 2, n = 0$ scalar QNMs with the WKB method using negative values of $\alpha$.} \label{tqnm1}
\end{center}

\begin{center}
\begin{tabular}{|c|c|c|c|c|} \hline
& $\alpha =0$ & $\alpha =0.1$ & $\alpha =0.2$ & $\alpha =0.3$ \\ \hline
$q=0.3$ & $0.53876-0.08927i$ & $0.53307-0.089182i$ & $0.53872-0.089248i$ & $%
0.53870-0.089234i$ \\ 
$0.4$ & $0.54524-0.08983i$ & $0.54521-0.089809i$ & $0.54518-0.089781i$ & $%
0.54514-0.089753i$ \\ 
$0.5$ & $0.55412-0.09055i$ & $0.55407-0.090504i$ & $0.55401-0.090452i$ & $%
0.55395-0.090400i$ \\ 
$0.6$ & $0.56593-0.09141i$ & $0.56584-0.091319i$ & $0.56575-0.091225i$ & $%
0.56565-0.091131i$ \\ 
$0.7$ & $0.58153-0.09234i$ & $0.58138-0.092175i$ & $0.58123-0.092002i$ & $%
0.58107-0.091826i$ \\ 
$0.8$ & $0.60240-0.09317i$ & $0.60216-0.092833i$ & $0.60190-0.092484i$ & $%
0.60162-0.092128i$ \\ 
$0.9$ & $0.63140-0.09322i$ & $0.63095-0.092412i$ & $0.63039-0.091557i$ & $%
0.62970-0.090671i$ \\ 
$1$ & $0.67495-0.08874i$ & $0.67303-0.085410i$ & $0.66936-0.081398i$ & $%
0.66178-0.077021i$ \\ \hline
\end{tabular} 
\captionof{table}{Variation of $l = 2, n = 0$ scalar QNMs with the WKB method using positive values of $\alpha$.} \label{tqnm2}
\end{center}
 In QNMs, there are two main components: the real part, which represents the oscillation's actual frequency, and the imaginary part, which is linked to the damping timescale and can be used to investigate the black hole's stability.  In our analysis, we studied the basic mode $l = 2$ and $n = 0$, where monopoles satisfy $(n<l)$ and the parameters $(q, \alpha)$ vary. $\alpha = 0$ corresponds to the original RN black hole. We found that increasing the electric charge parameter causes an increase in the real part of the frequencies as shown in Figure \ref{sh12}. However, $Im(\omega)$ shows a more dynamic change. The values drop until roughly $q\approx 0.85$, at which point a rising trend can be detected.\\ Figure \ref{sh13}
depicts the relationship between the quasinormal mode frequencies and the coupling parameter $\alpha$. When $q$ is fixed, we see that the real component steadily diminishes as $\alpha$ increases, particularly for large values of $q$. While the imaginary part steadily grows as $\alpha$ increases. However, the rise in the imaginary component is greater for higher values of $q$.

\begin{figure}
    \centering
    \includegraphics[scale=0.67]{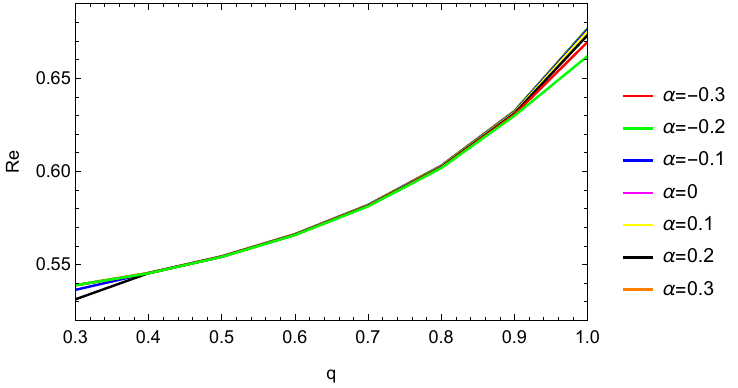}
\includegraphics[scale=0.5]{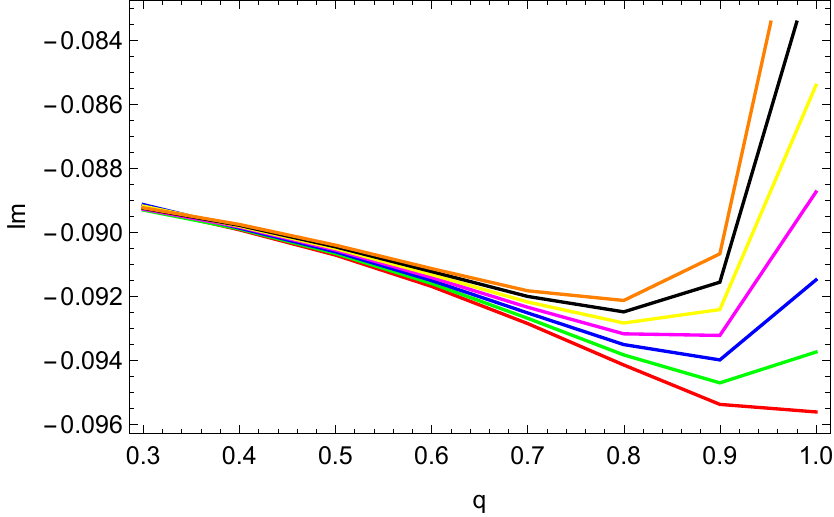}
    \caption{Variation of $n = 0, l = 2$ scalar QNMs with $q$ and $\alpha$. }
    \label{sh12}
\end{figure}
\begin{figure}
    \centering
    \includegraphics[scale=0.67]{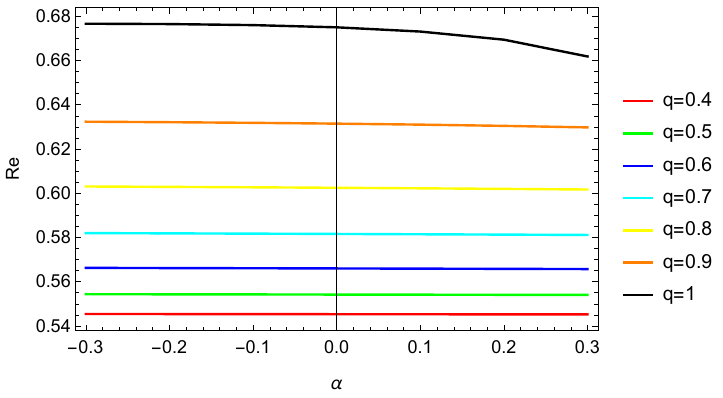}
\includegraphics[scale=0.5]{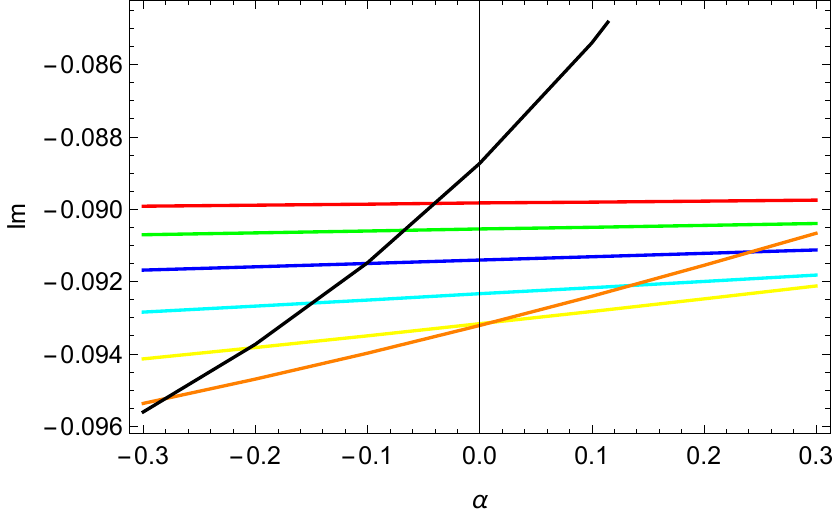}
    \caption{Variation of $n = 0, l = 2$ scalar QNMs with $q$ and $\alpha$. }
    \label{sh13}
\end{figure}

\section{Deflection angle of SCWBH via G\lowercase{s}B\lowercase{n}T\lowercase{h}} \label{sec6}

This section aims to clarify the method for calculating the deflection angle \cite{isPerlick:2021aok} within the context of gravitational theory that incorporates Weyl corrections. The computation assumes a weak field approximation and pertains to a massive SCWBH. The analysis is performed in a non-plasma medium, employing the Gauss-Bonnet theorem (GsBnTh) \cite{isGibbons:2008rj,isHalla:2020hee,isIslam:2020xmy,isOvgun:2018fte}. The GsBnTh plays a crucial role in connecting the intrinsic geometry of the metric to the topology of the region $\Xi_{\cal R}$, defined by its boundary $\partial \Xi_ {\cal R}$. This significant linkage is expressed concisely by the following equation \cite{isGibbons:2008rj}:
\begin{equation}\label{isGaussBonnet}
\iint_{\Xi_{\cal R}} \mathcal{K} \, dS + \oint_{\partial \Xi_{\cal R}} \mathfrak{k} \, dt + \sum_{z} \alpha_z = 2\pi \chi(\Xi_{\cal R}),
\end{equation}
where $\Xi_{\cal R} \subset S_{2D}$ denotes a regular region on a simple two-dimensional surface $\mathcal{A}_{sf}$, bounded by a closed, regular, and positively oriented curve $\partial  \Xi_{\cal R}$. Here, $\mathfrak{k}$ signifies the geodesic curvature of $\partial \Xi_{\cal R}$, defined as $\mathfrak{k} = \bar{g}(\nabla_{\dot{\gamma}} \dot{\gamma}, \ddot{\gamma})$, where the condition $\bar{g}(\dot{\gamma}, \dot{\gamma}) = 1$ holds and $\ddot{\gamma}$ is the unit acceleration vector. The term $\alpha_z$ at the $z^{th}$ vertex represents the exterior angle. In Eq. \eqref{isGaussBonnet}, $\chi(\Xi_{\cal R})$ denotes the Euler characteristic number \cite{isMa:2003uj}, and $\mathcal{K}$ denotes the Gaussian optical curvature \cite{isWeinberg:2013agg}, which is determined from the null geodesics influenced by the black hole \cite{isFernando:2012ue}. Given that light follows null geodesics (i.e., $ds^2=0$), these are meticulously selected to define the optical metric, which encapsulates the Riemannian geometry perceived by light. By applying the null condition along with $\theta = \pi /2$ (signifying the equatorial plane in the optical metric as a surface of revolution), we establish the following optical metric in a revised coordinate system:
\begin{equation}\label{is4}
dt^2=\tilde{g}_{ij} d\mathrm{x}^i d\mathrm{x}^j= d\varrho ^2+\digamma^2(\varrho)d\phi^2,
\end{equation}
where 
\begin{equation} \label{is4n}
\digamma(\varrho(r))=\sqrt{\frac{g(r)}{f(r)}},
\end{equation}
and $\varrho$ is named as the tortoise coordinate \cite{isNakajima:2021yfz,isTokgoz:2018irr,isPour} and it is given by
\begin{multline}
\varrho=\int \frac{1}{f(r)} \, dr \approx r + \left(M - \frac{\sqrt{M - q}\sqrt{M + q}(-2M^2 + q^2)}{2(-M^2 + q^2)}\right) \times \\
\ln\left(r + \frac{2M^2\left(M - \frac{\sqrt{M - q}\sqrt{M + q}(-2M^2 + q^2)}{2(-M^2 + q^2)}\right) + Mq^2 - 2q^2\left(M - \frac{\sqrt{M - q}\sqrt{M + q}(-2M^2 + q^2)}{2(-M^2 + q^2)}\right)}{-2M^2 + q^2}\right) \\
+ \left(M + \frac{\sqrt{M - q}\sqrt{M + q}(-2M^2 + q^2)}{2(-M^2 + q^2)}\right) \times \\
\ln\left(r + \frac{2M^2\left(M + \frac{\sqrt{M - q}\sqrt{M + q}(-2M^2 + q^2)}{2(-M^2 + q^2)}\right) + Mq^2 - 2q^2\left(M + \frac{\sqrt{M - q}\sqrt{M + q}(-2M^2 + q^2)}{2(-M^2 + q^2)}\right)}{-2M^2 + q^2}\right). \label{tort}
\end{multline}
It is worth noting that the metric function $f(r)$ involves nonlinear terms with high-order powers in the denominator, which can lead to a challenging integral that might not have a straightforward analytical solution. Therefore, Eq. \eqref{tort} has been given in approximate form. On the other hand, the non-vanishing Christoffel symbols \cite{isJusufi:2018kmk} associated with metric are computed as:
\begin{eqnarray} \label{is7.1}
\Gamma_{\phi \phi}^{\varrho}&=&-\digamma(\varrho)\frac{\mathrm{d}\digamma(\varrho)}{\mathrm{d}{\varrho}},\\
\Gamma_{\varrho\phi}^{\phi}&=&\frac{1}{\digamma(\varrho)}\frac{\mathrm{d}\digamma(\varrho)}{\mathrm{d}{\varrho}}. \label{is7.2}
\end{eqnarray}
Note that the determinant can be found as $\det\tilde{g}_{ij}=\digamma^{2}(\varrho)$. Consequently, one can evaluate the Gaussian optical curvature $\mathcal{K}$ \cite{isMandal:2023eae} as follows:
\begin{equation} \label{isK}
\mathcal{K}=-\frac{R_{\varrho \phi \varrho \phi}}{\det\tilde{g}_{r\phi}}=-\frac{1}{\digamma(\varrho)}\frac{\mathrm{d}^{2}\digamma(\varrho)}{\mathrm{d}{\varrho}^{2}}.
\end{equation}

The expression for the optical curvature $\mathcal{K}$ can alternatively be reformulated using the variable $r$ \cite{isOvgun:2018ran}. Consequently, one finds out the optical curvature as
\begin{eqnarray}
\mathcal{K} & = &-\frac{1}{\digamma(\varrho(r))}\left[f(r)\frac{df(r)}{\mathrm{d}r}\frac{\mathrm{d}\digamma(\varrho(r))}{\mathrm{d}r}+(f(r))^{2}\frac{\mathrm{d}^{2}\digamma(\varrho(r))}{\mathrm{d}r^{2}}\right].\label{is8}
\end{eqnarray}
After substituting Eqs. \eqref{equ2}, \eqref{egu}, and \eqref{is4n} in Eq. \eqref{is8}, the optical curvature appearing in a lengthy form can be given by
\begin{multline}
\mathcal{K} = \frac{r^{-14}}{2025{(9r^4 + 4\alpha q^2}{r^2})^2}\bigg( 230400 r^4 \alpha^4 q^8 - 328050 Mr^{19} - 2112000 Mr^3 \alpha^4 q^8 \\
- 984150 r^{17} Mq^2
- 10366380r^{14} q^4 \alpha - 6969240 r^{12} q^6 \alpha +
3933360 r^{10} q^6 \alpha^2 + 6469632 r^8 q^8 \alpha^2\\
+ 4351680 \alpha^3 q^8 r^6 + 5960448 \alpha^3 q^{10} r^4 + 1497600 \alpha^4 q^{10} r^2 - 
2624400 r^{16} \alpha q^2 + 907200 r^{12} \alpha^2 q^4 \\
+ 518400 r^8 \alpha^3 q^6+ 492075 r^{18} M^2 + 328050 r^{16} q^4 + 492075 r^{18} q^2 + 1730560 \alpha^4 q^{12}\\
+ 16329600 r^{15} \alpha q^2 M
- 20995200 r^{14} \alpha q^2 M^2 + 24756840 r^{13} \alpha q^4 M\\
- 7387200 r^{11} \alpha^2 q^4 M 
+18468000 r^{10} \alpha^2 q^4 M^2 - 22019040 r^9 \alpha^2 q^6 M \\
- 5760000 r^7 \alpha^3 q^6 M + 13824000 r^6 \alpha^3 q^6 M^2 - 18921600 r^5 \alpha^3 q^8 M + 4320000 \alpha^4 q^8 M^2 r^2\\
- 5657600 \alpha^4 q^{10} Mr \bigg) 
\end{multline}
One can observe that the Gaussian optical curvature depends on various parameters such as mass $M$, $q$, and $\alpha$. Now, we would like to determine the deflection angle employing the GsBnTh. Now, one can consider very large radial distance $r \equiv R \rightarrow \infty$, in such a limit, the jump angle converges to $\pi/2$, ensuring $\theta_{0}+\theta_{S}=\pi$, and the Euler characteristic number becomes unity  \cite{isAtamurotov:2021cgh}. As a consequence, Eq. \eqref{isGaussBonnet} recasts in

\begin{equation}\label{is43a}
\iint_{\Xi_{\cal R}} \mathcal{K} d S+\oint_{\partial \Xi_{\cal R}} \mathfrak{k} d t+\alpha_{z}=2 \pi  {\chi}\left(\Xi_{\cal {R}}\right),
\end{equation}
where $\alpha_{z}$ represents the total angle of jumps, set as $\pi$, and as $\mathcal{A}_{sf}\rightarrow\infty$, the geodesic curvature $\mathfrak{k}\left(C_{\cal R}\right)$ can be expressed as the magnitude of the gradient of $\dot{C}_{\cal R}$ with respect to $\dot{C}_{\cal R}$ itself: $\left|\nabla_{\dot{C}_{\mathcal{R}}} \dot{C}_{\mathcal{R}}\right|$. The radial component of geodesic curvature can then be obtained as \cite{isOkyay:2021nnh}:

\begin{equation}\label{is44}
\left(\nabla_{\dot{C}_{\cal R}} \dot{C}_{\cal R}\right)^{r}=\dot{C}_{\cal R}^{\phi} \partial_{\varphi} \dot{C}_{\cal R}^{r}+\Gamma_{\phi \phi}^{r_{*}}\left(\dot{C}_{\cal R}^{\phi}\right)^{2}.
\end{equation}

For very large distance ${\cal R}$, where $C_{\cal R}:=r(\phi)={\cal R}$ is constant, one can get $\left(\dot{C}_{R}^{\phi}\right)^{2}=\digamma(\varrho(r))$. After making some algebra, the geodesic curvature can then be obtained as follows \cite{isGibbons:2008rj}:
\begin{equation}\label{is45}
\left(\nabla_{\dot{C}_{\cal R}^{r}} \dot{C}_{{\cal R}}^{r}\right)^{r} \rightarrow \frac{1}{{\cal R}},
\end{equation}
which means that $\mathfrak{k}\left(C_{\mathcal{R}}\right) \rightarrow \frac{1}{\mathcal{R}}$. Utilizing the optical metric \eqref{is4}, we get $dt={\cal R} d \phi$ leading to the following expression:

\begin{equation}\label{is46}
\mathfrak{k}(C_{\cal R})dt=\lim_{{\cal R}\to\infty}[\mathfrak{k}(C_{\cal R})dt]
         =\lim_{{\cal R}\to\infty}\left[\sqrt{\frac{ \tilde{g}^{\phi\phi}}{4\tilde{g}_{r_{*}r_{*}} }}\left(\frac{\partial\tilde{g}_{\phi\phi}}{\partial r_{*}}\right)\right]d\phi
         =d\phi.
\end{equation}

Taking into account all of the earlier findings, the GsBnTh results in
\begin{equation}\label{is47}
\iint_{\Xi_{{\cal R}}} \mathcal{K} d S_{rf}+\oint_{\partial \Xi_{{\cal R}}} \mathfrak{k} d t\stackrel{{\cal R}\rightarrow \infty} {=}  \iint_{S_{rf:\infty}} \mathcal{K} d \mathcal{A}_{sf}+\int_{0}^{\pi+\tilde{\delta}} d \phi.
\end{equation}
\begin{figure*}
    \centering
{\includegraphics[scale=0.6]{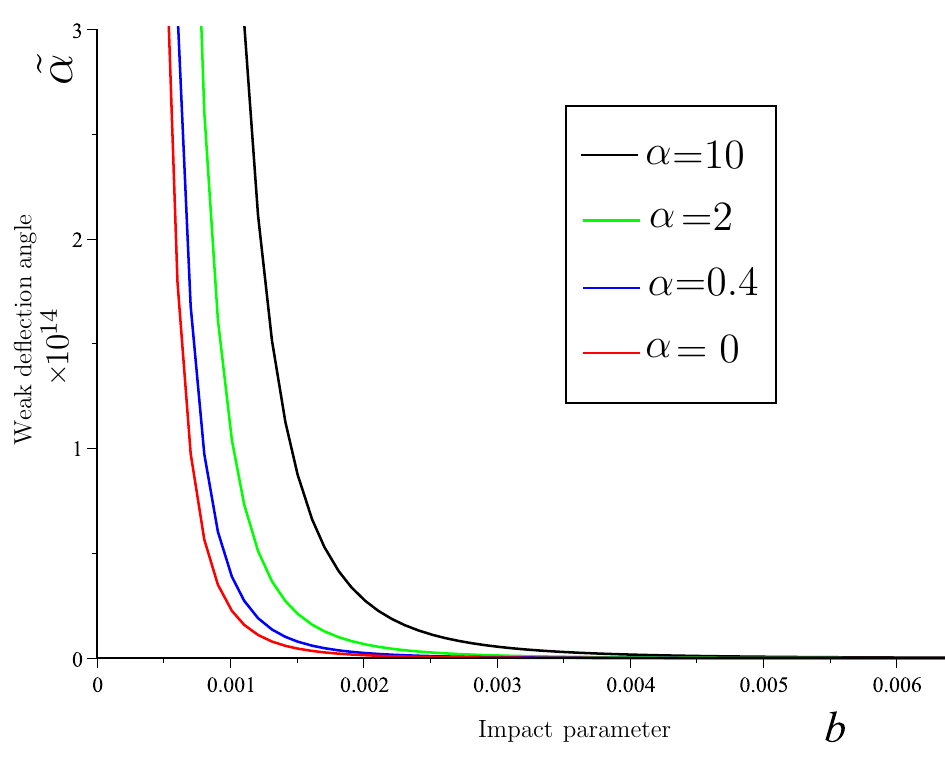} }\qquad
    {\includegraphics[scale=0.6]{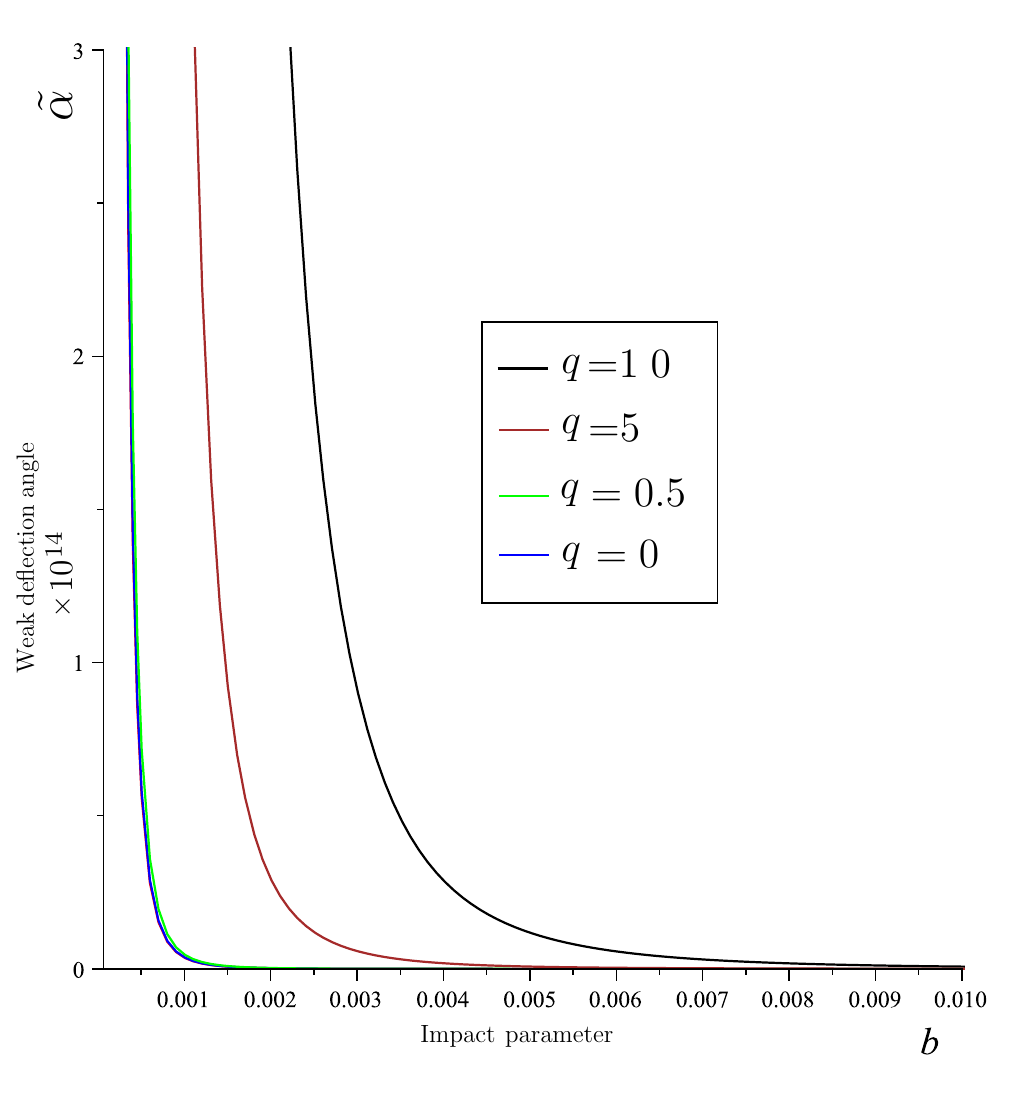}}
    \caption{The profile of the deflection angle around the SCWBH as a function of impact parameter $\mathfrak{p}$ for different values of Weyl correction parameter $\alpha$ (up panel, $q=0.3$) and charge $q$ (down panel, $\alpha=1$), respectively, for fixed $M=1$. }
    \label{fig:def_an}
\end{figure*}

In the weak-deflection limit scenario, the trajectory of a light ray is simplified to a straight line, represented by the equation $r(t)\equiv\mathfrak{B}=\frac{b}{sin {\phi}}$, where $b$ denotes the impact parameter \cite{isGibbons:2008rj}. Utilizing this expression, one can compute the deflection angle $\tilde{\alpha}$ as follows \cite{isMandal:2023eae}:

\begin{equation}\label{is48}
\tilde{\alpha}=-\int_{0}^{\pi} \int_{\mathfrak{B}}^{\infty} \mathcal{K} \mathcal{A}_{sf}=-\int_{0}^{\pi} \int_{\mathfrak{B}}^{\infty}\frac{\mathcal{K} \sqrt{\operatorname{det} \tilde{g}}}{F(r)} dr d \phi =-\int_{0}^{\pi} \int_{\mathfrak{B}}^{\infty}  \frac{g(r)\mathcal{K}}{f(r)^\frac{3}{2}} d r d \phi.
\end{equation}
Therefore, in a non-plasma medium, for the spacetime of a spherically symmetric statically charged black hole within a gravitational field coupled with Weyl corrections, the deflection angle \big(by using Eqs. \eqref{equ2}, \eqref{egu}, \eqref{tort}, and \eqref{isK} in Eq. \eqref{is48}\big) reads
\begin{equation}\label{is49}
\tilde{\alpha}=-\int_{0}^{\pi} \int_{\mathfrak{B}}^{\infty}  \bigg(-\frac{45 \sqrt{9 r^4+4 \alpha q^2} \sqrt{5} r^2\left(-2 M r^3+3 M^2 r^2+3 q^2 r^2-6 q^2 M r-16 \alpha q^2+2 q^4\right)}{\left(45 r^6-90 M r^5+45 q^2 r^4-60 \alpha q^2 r^2+200 \alpha q^2 M r-104 \alpha q^4\right)^{(3 / 2)}}\bigg)drd\phi,
\end{equation}
which can be approximated to the following form:
\begin{equation}\label{is49}
\tilde{\alpha}\approx-\int_{0}^{\pi} \int_{\mathfrak{B}}^{\infty} \left(\frac{2 M}{r^2}+\frac{3 M^2}{r^3}-\frac{3 q^2}{r^3}+\frac{6 M^3}{r^4}-\frac{6 M q^2}{r^4}-\frac{15 M^2 q^2}{r^5}+\frac{25 M^4}{2 r^5}+\frac{5 q^4}{2 r^5}+\frac{16 \alpha q^2}{r^5}\right)drd\phi.
\end{equation}
After making straightforward calculations, one can obtain the following deflection angle:
\begin{equation} \label{is50}
\tilde{\alpha}\approx -\frac{3 \pi q^2}{4 b^2}+\frac{3 \pi q^2 \alpha}{2 b^4}+\frac{15 \pi q^4}{64 b^4}+\left(\frac{4}{b}-\frac{8 q^2}{3 b^3}\right) M+\left(\frac{3 \pi}{4 b^2}-\frac{45 \pi q^2}{32 b^4}\right) M^2+\frac{8 M^3}{3 b^3}+\frac{75 \pi M^4}{64 b^4}.
\end{equation}

Thus, one can immediately observe how the Weyl corrections ($\alpha$) and the charge $q$ influence the deflection angle (\ref{is50}), which reduces to the RN black hole case \cite{isJaved:2023iih,isJusufi:2015laa,isPang:2018jpm} in the absence of $\alpha=0$. At this point, it is important to highlight that the method of GsBnTh can be applied to any asymptotically flat Riemannian optical metric due to its distinctive topological characteristics. \\
We then analyze the remarkable effect of the Weyl correction parameter $\alpha$ with black hole charge $q$ on the weak deflection angle $\tilde{\alpha}$ by adapting the method of GsBnTh method. To this end, we show the dependence of the deflection angle profile of the impact parameter $b$ in Fig. \ref{fig:def_an}. For being more informative, in Fig. \ref{fig:def_an}, the top panel reflects the impact of the Weyl correction parameter $\alpha$ with fixed charge $M,q$ on the profile of the deflection angle, while the bottom panel reflects the impact of black hole charge $q$ on fixed $M,\alpha$. From Fig. \ref{fig:def_an}, both the Weyl  correction parameter $\alpha$ and the charge $q$ affect the deflection angle $\tilde{\alpha}$ in such a way that they increase the (weak) deflection angle with their increasing values. In other words, $\alpha$ and $q$ exhibit analogous physical influences, leading to a proportional effect on the deflection angle.

\section{Conclusion} \label{sec7}
The study presented herein extensively investigates the observable properties of the SCWBH, which was obtained within the framework of Einstein-Maxwell action extended by Weyl tensor corrections. Theoretical predictions concerning these modified black holes were explored, focusing on their impact on surrounding spacetime properties, including geodesics, photon spheres, quasinormal modes, and the bending of light.

One of the significant findings of this study is the modification of geodesic paths around the black holes. The analysis revealed that the Weyl corrections introduce deviations in both particle and photon trajectories, which could influence the dynamics of objects in the vicinity of these black holes. These results were substantiated by examining the effective potential $V_{\text{eff}}$ (\ref{effp88}). The modifications in the geodesic structure were analyzed, highlighting changes in the stability conditions and the nature of orbits. It is found that increasing the coupling parameter causes the ISCO radius to
slightly rise. Furthermore, the study delved into the properties of photon spheres and black hole shadows. The radius and visibility of photon spheres, which are critical in determining the observable features of black holes slightly shrinks in the presence of Weyl corrections. The SCWBH's shadow radius $R_{\text{sh}}$ was observed to change depending on the Weyl tensor coupling strength, demonstrating the possibility of detecting these phenomena using modern astronomical methods.

The study of quasinormal modes offered understanding into the stability of the spacetime around these black holes. The results showed that the real component decreases with increasing $\alpha$, especially for high $q$ values. The imaginary part grows steadily as $\alpha$ increases. These modifications are represented in the frequency variations $\delta \omega$, which can be described as a function of the Weyl correction parameter $\alpha$. Additionally, the bending of light by these black holes was examined using the GsBnTh, leading to a detailed expression for the deflection angle $\tilde{\alpha}$, given by Eq. \eqref{is50}. The results underscored the impact of these corrections on gravitational lensing, potentially allowing for novel tests of gravity theories.

In conclusion, the integration of Weyl tensor corrections into the Einstein-Maxwell action for black holes provides a fertile ground for theoretical predictions and observational confirmations. The modified gravitational dynamics, as revealed through changes in geodesic behavior, photon sphere properties, quasinormal modes, and light deflection, offer compelling evidence that such corrections could be integral to a more comprehensive understanding of gravity. Future observational efforts, supported by advancements in astronomical techniques, could provide critical data to validate or refine these theoretical predictions, contributing to the broader quest for a unified theory of gravity.

\section*{Acknowledgments}
{\color{black}\.{I}.S thanks for the contributions of T\"{U}B\.{I}TAK, ANKOS, and SCOAP3. He also acknowledges the networking support of COST Actions CA18108, CA22113, CA21106, and CA23130.}
\\ \large{Conflict of interest}\\ The authors declare no conflict of interest. \\
\large{Data availability}\\
No data was used for the research described in the article.

\end{document}